\documentclass[11pt]{article}

\usepackage{fullpage}
\usepackage{times,graphicx,latexsym}
\usepackage{amsmath}
\usepackage{url}
\usepackage{booktabs}
\usepackage{subfig}
\usepackage{mymath}

\usepackage{tikz}
\usepackage{url}
\usepackage{algorithm,algorithmic}
\usepackage{authblk}

\graphicspath{{./}}
\usetikzlibrary{arrows,positioning,matrix}

\title{Cumulative Revision Map}

\author{Seungyeon Kim\textsuperscript{*}}
\author{Joshua V. Dillon}
\author{Guy Lebanon}
\affil{
  College of Computing\\
  Georgia Institute of Technology\\
  Atlanta, Georgia
}

\begin{document}
\date{\today}
\maketitle
\thispagestyle{empty}
\let\oldthefootnote\thefootnote
\renewcommand{\thefootnote}{\fnsymbol{footnote}}
\footnotetext[1]{Email: \url{seungyeon.kim@gatech.edu}}
\let\thefootnote\oldthefootnote

\maketitle

\begin{abstract}
  Unlike static documents, version-controlled documents are edited by one or more authors over a certain period of time. Examples include large scale computer code, papers authored by a team of scientists, and online discussion boards. Such collaborative revision process makes traditional document modeling and visualization techniques inappropriate. In this paper we propose a new visualization technique for version-controlled documents that reveals interesting authoring patterns in papers, computer code and Wikipedia articles. The revealed authoring patterns are useful for the readers, participants in the authoring process, and supervisors.
\end{abstract}

\section{Introduction} \label{sec:intro}

Version-controlled documents are usually authored by several users and updated over a certain period of time unlike static documents. One instance of version-controlled documents are large software projects developed by teams of software engineers over a period of several weeks. Another example of version-controlled documents are scientific papers written by teams of scientists across multiple geographic location. A third instance of such documents is online discussion boards like Slashdot or Google Wave. In each case the authoring process is composed of a sequence of document revisions annotated with the date, the identity of the author, and occasionally revision comments. 

The importance of such collaboratively authored documents has recently increased substantially with the availability of collaborative productivity tools such as Subversion/GIT, Google Docs, MS Office 2010, and online forums like Wikipedia and Wordpress. These tools or websites maintain a complete revision history which may be used to recreate the entire authoring process (rather than just the final document). 

Visualizing version-controlled documents has a number of important applications. It may assist authors or code developers in determining what is the current project status and what they should work on next (either revise or avoid revising). It may assist managers in ensuring that the code or document development process progresses adequately, and if not identify the problem. For example, are there authoring inefficiencies such as certain authors consistently overwriting their colleagues. It may also be used to expose collaborative authoring patters leading to disinformation, which is a serious problem in the Wikipedia project. 

In this paper we develop a representation, called cumulative revision map (CRM), for visualizing version-controlled documents. Most well known document visualization techniques are appropriate only for static documents, but they are unable to capture patterns in the authoring process. Our approach displays the entire revision history as a two dimensional diagram where rows correspond to revisions and columns to document position. The spatial arrangement of additions and deletions reveals informative authoring patterns. After explaining the CRM visualization method we demonstrate it on a combination of synthetic documents and real world documents including Wikipedia articles, scientific papers, and computer code. We also compare the CRM visualization technique to related work such as History Flow and discuss their pros and cons. 

\section{Related Work}
Several attempts have been made to visualize themes and topics in documents, either by keeping track of the word distribution or by dimensionality reduction techniques e.g.,\cite{Fortuna2005,Havre2002,Spoerri1993,nvacbook}. Such studies tend to visualize a corpus of unrelated documents as opposed to ordered collections of revisions which we explore.

Document visualization has gained considerable real world and research interest due to the inherent complexity of text and the overwhelming extent of digital text archives such as the Internet.   Collections of version-controlled documents, such as code repositories and Google docs, compound these challenges by storing documents as they evolve over time and by several authors. Techniques for visualizing version-controlled documents tend to focus more on temporal and collaborative aspects and less on content.

A partial list of references for text visualization are \cite{Spoerri1993,Havre2001,Havre2002,Fortuna2005,Viegas2006,Blei2006} with additional references available in \cite{Thomas2005}. A selection of software systems for visualizing text corpora are IN-SPIRE\footnote{\url{http://in-spire.pnl.gov}}, Jigsaw\footnote{\url{http://www.cc.gatech.edu/gvu/ii/jigsaw/}}, Enron corpus viewer\footnote{\url{http://jheer.org/enron}}, Thomson's refviz\footnote{\url{http://www.refviz.com}}, and the Science topic browser\footnote{\url{http://www.cs.cmu.edu/~lemur/science/}}

\subsection{Visualizing Word Histograms}

Visualizing numeric data, such as word histograms, serves a foundational role in visualizing complicated textual objects.  Monographs describing traditional visualization techniques are \cite{Cleveland1993,Tufte2001} while less traditional approaches for visual data exploration are surveyed in \cite{Oliveira2003}. Some interesting ideas concerning visualizing low-dimensional numeric time series are \cite{Weber2001,Hochheiser2004}. Recent trends in the area of time series visualization are mostly concerned with interactive visualization and with multiple or vector-valued time series. An interesting exposition of the state-of-the-art and future vision in the related field of visual analytics is \cite{Thomas2005}.

The use of $n$-grams to convert categorical sequences to numeric vectors is used extensively in the fields of information retrieval, speech recognition, and natural language processing.  Recent monographs describing the use of $n$-grams in these areas are \cite{Jelinek1999,Manning1999,Yates1999}. Visualizing $n$-grams is usually accomplished through statistical dimensionality reduction techniques.  Methods such as principal component analysis and multidimensional scaling are surveyed in \cite{Fodor2002} while \cite{Lee2007} reviews non-linear techniques for dimensionality reduction.

\subsection{Visualizing Version-controlled Documents}
Visualizations for version-controlled documents primarily focus on programming (code) repositories rather than more traditional documents.  Although traditional document authoring is fundamentally different in style and content, one could imagine using these techniques for depicting the life-cycle of general documents.

History Flow and Text-animated-transition \cite{Viegas2004,Chevalier2010} are outstanding techniques fir visualizing the authoring process in a single document. This is also the approach we take in our paper. Other approaches attempt to visualize an entire repository containing a large number of documents. The resulting visualization focuses on which files are edited by whom and when rather than how the text content changes. 

Good overviews of techniques for visualizing the software evolution process are \cite{Storey2005,Diehl2007}. Specific examples include SeeSoft \cite{Eick1992}, a line-by-line visualization of source code, as well as Augur \cite{Froehlich2004} and Advizor \cite{Eick2002}.  The latter two are collections of visualizations, such as 2D and 2.5D matrix views which identify file and source changes in terms of project branch, date, author, etc. Cenqua Fisheye\footnote{\url{http://fisheye.cenqua.com/}} is one such commercial tool for visually interacting with software repositories, however the interface consists of text-centric and graphical displays displaying line charts and histograms. The StarGate project \cite{Ogawa2008} and CodeSaw\footnote{\url{http://social.cs.uiuc.edu/projects/codesaw.html}} serve roles similar to FishEye, i.e. tracking where and to what extent authors are concentrating their efforts, but provide a less static presentation.

An increasing number of recent visualization techniques emphasize aesthetics and result in a more qualitative rather than quantitative presentations of information. Organic visualizations \cite{Fry2000} use non-standard visual mechanisms, such as swirling clouds and blooming flowers, in which data members interact to exhibit emergent structure. Additional examples include gource\footnote{\url{http://code.google.com/p/gource/}} and code\_swarm \cite{Ogawa2009}. 

Most of the tools above are primarily intended to provide an understanding of the evolution of a collection of documents.  Conversely, we present techniques for visually exploring the life-cycle of a single one document. Obviously both serve fundamentally different roles and answer different questions. The former presents a high level overview picture of the evolution of a document repository while the latter provides more detailed view concerning the authoring process of a specific document.

\section{Cumulative Revision Map}

In this section, we introduce the Cumulative Revision Map and the precise techniques used to generate the visualization. We provide additional arguments for our design choices by contrasting with the Unix tool ``diff'' and History Flow, the visualization techniques most similar to this work.

\subsection{Data Reduction Principles}\label{sec:reduction}
``Diff'' has been a mainstay of the Unix system since its inception in the 70s and arguably remains the most useful general-purpose revision visualization system.  Diff input consists of two files, a reference document and a proposal document, and outputs a sequence of line by line edits, i.e., add and delete.  Such edits, if applied to the reference file, would exactly yield the proposal file. Since there are an infinite number of possible edit sequences, diff solves the longest common subsequence (LCS) problem to characterize a sense of minimal change needed to realize the proposal file.

The defining characteristic of diff is its conveyance of only the differences between two documents.  We abstractly refer to this difference as the document-document delta.  By presenting only the delta, the amount of data the user must interpret is reduced, often significantly, and allows him or her to form a mental picture of the change between any two revisions and how this change is correlated with other events, such as authorship, content, or time.  Such an approach has the advantage of being lossless, given the reference document.

In principle, one could examine multiple revisions through repeated application of diff.  Unfortunately, such an approach is ill-suited for many modern collaborative settings where there may be hundreds or thousands of revisions over the course of months, years, or even decades.  In these cases, the complexity of data is characterized not just by the delta, but by the number of such deltas. A sequence of deltas rapidly becomes overly complicated and belies the data-reduction principles motivating the use of diff in the first place. As a consequence, the user's ability to make high-level characterizations of the document's evolution is compromised--a task vitally important in settings where there may be tens or hundreds of authors (e.g., Wikipedia).

The CRM fundamentally draws upon the same data reduction principles of diff, but extends this delta-type reduction to the time domain.  As with diff, CRM depicts only the changes (between documents) but unlike diff, the characterization is graphical rather than textual. Through this presentation, the CRM is capable of representing 100s or 1000s of revisions while simultaneously representing the changes to the document in their entirety.

We make these notions precise by characterizing a revision delta as a 4-tuple, denoted $\Delta=(\mathcal{E},\mathcal{P},\mathcal{X},\mathcal{Y})$. Here $\mathcal{E}=\{\text{delete},\text{add},\text{delete}\cdot\text{add}\}$ denotes the possible edit operations, $\mathcal{P}=\{\text{``string''}\}$ denotes the edit payload (possibly empty), $\mathcal{X}$ the position in document, and $\mathcal{Y}$ the revision number.  Denoting an instance of $\Delta$ as $\delta$ one can encode the life-cycle of any document as a sequence of delta instances, $(\delta_i)_{i\in I}$.  For example, adding a LaTeX section header after the $271$-th token of revision $8$ is denoted as \[(\text{add}, ``\verb|\section{Introduction}|",271,8).\] More generally a delta could also contain meta-information such as author or IP, however we omit this for simplicity.

\subsection{Schema}

As motivated by Sec.~\ref{sec:reduction}, the life-cycle of a document can be efficiently encoded as a sequence of revision deltas $(\delta_i)_{i\in I}$ each an instantiation of the 4-tuple $\Delta=(\mathcal{E},\mathcal{P},\mathcal{X},\mathcal{Y})$ with members corresponding to an edit operation, payload, position, and time (respectively).

The cumulative revision map visualizes this delta sequence as color-coded elements of a matrix indexed by both position in document and revision number.  We refer to these indices as \textit{space-time coordinates} where spatial position refers to a particular position in the document (e.g., word count, line number, or byte) and temporal position refers to a particular revision number.  For example, Figure~\ref{fig:crm1} (bottom) depicts the revision history of an actual conference paper.  The CRM $x$-axis represents space with the left-most column representing the document beginning and the right-most the document end.  Time is characterized by the $y$-axis with each row indicating a subsequent revision; the top-most row corresponds to the first revision and the bottom-most row the most recent revision.

The basic edit operations are coded by color; adds are gray, deletes are red.  Since the CRM graphically depicts deltas by color-coded space-time coordinates, only the payload information is not graphically depicted.  This information is conveyed interactively via a simple pop-up mechanism (Fig.~\ref{fig:interactive}) attached to each matrix element. The pastel, horizontal bands indicate which author made the changes and vertical bands correspond to sections (when such a construct exists in the document).  Tracing the faint vertical lines throughout shows the portions of the document which persist to the most recent revision.

The CRM can be interpreted in several different ways.  Much like its diff analog, a particular revision can be recovered by applying the edit operations (top-down) until the revision of interest is reached.  More generally, the visualization scheme naturally depicts a high-level characterization of how often the document was edited and with what locality.  Through the pop-up mechanism, users are able to obtain precise edit details (e.g., the actual diff output) by simply clicking on the element in question.

\subsection{Design}
CRM maintains entire addition and deletion history of a document. The history is maintained by a graph structure with each nodes containing a subsequence of a document. In each revisions, CRM solves a Longest Common Subsequence problem (like the \emph{unix diff}) between sequence at a revision and the previous one. LCS finds the minimal addition and deletion(revision delta) between two revisions; which is the edit history at certain revision. Using those revision deltas, CRM updates its graph preserving cumulatively. Each update will have those operations: unchanged part intact, splitting relevant node to add new content and to delete some parts of a node. The full algorithm is described in Algorithm~\ref{alg:crm}. 

\begin{algorithm}
  \caption{The process of Cumulative Revision Map}  \label{alg:crm}
  \begin{algorithmic}
    \STATE $G$ = empty graph
    \FOR{each revision}
      \STATE solve LCS problem to find additions and deletion
      \FOR{each edit operation}
        \STATE find relevant node in $G$
        \IF{addition}
          \IF{add at the beginning}
            \STATE make a new node with new content and make a link
          \ELSIF{add at the ending}
            \STATE make a new node with new content and make a link
          \ELSE
            \STATE split the node in the position of addition
            \STATE make a new node with new content
            \STATE make links to split nodes
          \ENDIF
        \ELSIF{deletion}
          \STATE split the node to separate the deleted part
          \STATE mark the node \emph{dead}
          \STATE attach the node to the other node
        \ENDIF
      \ENDFOR
    \ENDFOR
    \STATE layout the graph $G$
    \STATE draw cumulative-change bars
  \end{algorithmic}
\end{algorithm}

An illustrative example of the process of CRM is shown in Figure~\ref{fig:crmProcess}. An integer in the node is the contents of the node, which is a subsequence of document. Gray boxes indicates persisting parts and reds for non-persisting parts. The arrow connecting gray nodes means the sequential flow of the latest revision of the graph.

At revision 2, $6$ was added. CRM finds relevant node, the one with $\{1,2,3,4,5\}$ and split it to $\{1,2,3\}$ and $\{4,5\}$ to insert a new node $\{6\}$ in the right position. The updated graph will maintain the document at revision 2. We can confirm the contents of the document while following the gray nodes connected with arrow edges. Revision 3 has a delete operation. Node $\{1,2,3\}$ is split to mark the deleted subsequence $\{1\}$. Revision 4 has one deletion and one deletion. Both operation is done separately without interfering each other. As a result, the final CRM maintains its contents as well as its entire edit histories.

The vertical position of a node in CRM follows the added revision of the contents. The horizontal position of a node follows relative position for the latest revision. As a result, the top-bottom projection of persistent nodes result the document of latest revision. (Figure~\ref{fig:crmProjection})

\definecolor{lightred}{rgb}{0.8, 0.5, 0.5}
\definecolor{lightlightgray}{rgb}{0.85, 0.85, 0.85}

\tikzset{ 
        >=stealth',
        live/.style={
     rectangle,
     draw=black,
     fill=lightlightgray,
     text centered
   },
    dead/.style={
      rectangle,
      draw=black,
      fill=lightred,
      text centered
    },
    desc/.style={
      minimum height=2em,
    },
    rev/.style={
      rectangle,
    },
    nexte/.style={
      ->,
      thick,
    },
    deade/.style={
      -,
      thin
    }
}

\begin{figure} 
  \centering
  \begin{tikzpicture}[node distance=0.7cm, right]
    \matrix[row sep=0.3cm]{
    \node[desc](s) at (-0.2,0.6){rev 1 with (1,2,3,4,5) };
    \node[live](1) at (0,0){1 2 3 4 5};
    \node[rev] at (4.5,0) {(rev1)};\\
    \node[desc](s) at (-0.2,0.6) {rev 2 with (1,2,3,6,4,5), 6 added };
    \node[live](1) at (0,0){1 2 3};
    \node[right of=1](d){};
    \node[live,below of=d](2){6};
    \node[live,right of=d](3){4 5};
    \node[rev](r1) at (4.5,0){(rev1)};
    \node[rev, below of=r1]{(rev2)};
    \path[nexte, bend right=45] 
      (1) edge (2)
      (2) edge (3);\\
    \node[desc](s) at (-0.2,0.6){rev 3 with (2,3,6,4,5), 1 deleted};
    \node[dead](1) at (0,0){1};
    \node[live,right of=1](23){2 3};
    \node[right of=23](d){};
    \node[live,below of=d](6){6};
    \node[live,right of=d](4){4 5};
    \node[rev](r1) at (4.5,0){(rev1)};
    \node[rev, below of=r1](r2){(rev2)};
    \path[nexte, bend right=45] 
      (23) edge (6)
      (6) edge (4);
    \path[deade]
      (1) edge (23);\\
    \node[desc](s) at (-0.2,0.6){rev 4 with (2,7,3,6,5), 7 added and 4 deleted };
    \node[dead](1) at (0,0){1};
    \node[live,right of=1](2){2};
    \node[right of=2](d1){};
    \node[below of=d1](d2){};
    \node[below of=d2](d3){};
    \node[live,below of=d3](7){7};
    \node[live,right of=d1](3){3};
    \node[right of=3](d){};
    \node[live,below of=d](6){6};
    \node[dead,right of=3](4){4};
    \node[live,right of=4](5){5};
    \node[rev](r1) at (4.5,0){(rev1)};
    \node[rev, below of=r1](r2){(rev2)};
    \node[rev, below of=r2](r3){(rev3)};
    \node[rev, below of=r3](r4){(rev4)};
    \path[nexte, bend right=45] 
      (2) edge (7)
      (7) edge (3)
      (6) edge (5);
    \path[nexte]
      (3) edge (6);
    \path[deade]
      (1) edge (2)
      (4) edge (5);\\
    };
  \end{tikzpicture}
  \caption{Cumulative Revision Map process. (rev1) Started with a document (1,2,3,4,5). (rev2) The node was split to insert a new content. (rev3) Node `123' was split to mark the deleted node. (rev4) Node `45' split and new node `7' was inserted.}
  \label{fig:crmProcess}
\end{figure}
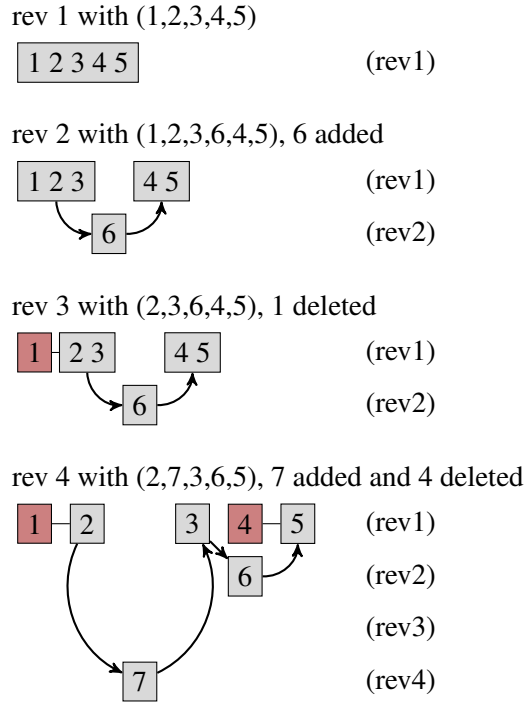

\begin{figure} 
  \centering
  \begin{tikzpicture}[node distance=0.7cm, right]
    \node[live](2){2};
    \node[live,right of=2](7){7};
    \node[live,right of=7](3){3};
    \node[live,right of=3](6){6};
    \node[live,right of=6](5){5};
    \path[nexte]
      (2) edge (7)
      (7) edge (3)
      (3) edge (6)
      (6) edge (5);
  \end{tikzpicture}
  \caption{The top-down projection of persistent nodes of revision 4 of the graph of Figure~\ref{fig:crmProcess}. It is exactly the same contents of the document at latest revision}
  \label{fig:crmProjection}
\end{figure}
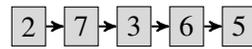

\subsection{Scalablity and Interactivity}
\begin{figure}
  \centering
  \includegraphics[width=.5\linewidth]{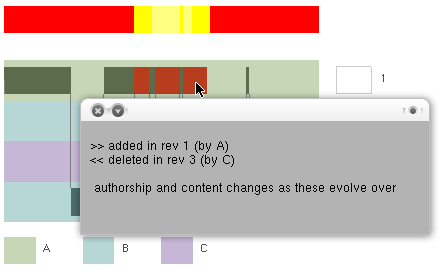}
  \caption{User interaction with CRM. User can select each node to see the contents and edit operations along with the revision number}
  \label{fig:interactive}
\end{figure}

CRM is scalable and interactive. The nodes and edges could be simplified to give a clear representation of large datasets. Edges could be changed to vertical lines while shrinking all horizontal gaps between nodes as shown in top section of Figure~\ref{fig:synthetic}. This simplification approach also gives more concise document location along horizontal layout. Moreover, user can intuitively pinpoint a node with mouse pointer to find out what was written and when the change was made. (Figure~\ref{fig:interactive})

\subsection{Comparison to Related Work}

CRM is distinct from other revision visualization systems in its visual characterization of a revised document.  By exploiting the sparsity present in a delta encoding of the document lifecycle, the graphical depiction preserves relevant information in a manner that is naturally sparse yet retains relevant information.  This sparsity also facilitates scalability, i.e., the entire document lifecycle can be visualized--a characteristic not present in many related works.

One preexisting work similar to the CRM is History Flow (HF) \cite{Viegas2004}. Like the CRM, HF represents a document's spatial characteristics (e.g., tokens) and temporal characteristics (e.g., revisions).  Although both CRM and History Flow depict revision data, the two approaches differ fundamentally.  CRM implicitly represents a document through visualizing only the deltas, while History Flow represents the document in its entirety at each revision.  Loosely speaking, one can think of the columns of HF as a snapshot of the document at a revision while the CRM can be regarded as a difference between snapshots.
	
Indeed, the CRM can be transformed into HF by accumulating the rows (and transposing the result). However, HF cannot quite be transformed into CRM.  The differences between HF columns would bare some semblance to the CRM, however, the natural sparsity in the CRM encoding allows the visualization of secondary information, e.g., edit persistence, section, etc., without overwhelming the visualization scheme.  To encode this data, the HF scheme relies on different visualization modalities through a user interface.

Maintaining changes has several advantages over maintaining all snapshots.  First, it is simpler. Without redundancy, the amount of information for a version-controlled document is smaller than its snapshot-based counterpart. Such space-efficiency is critical for version-controlled documents with long histories. Second, the CRM exploits whitespaces information in a manner that conveys information and yields a representation which easier to conceptualize.

\begin{figure*}[htp]
  \centering
  \begin{tabular}{cccc}
  \subfloat[Add at the beginning of the contents]{\includegraphics[scale=0.60,trim=1in 12.55in 5in 0in,clip=true]{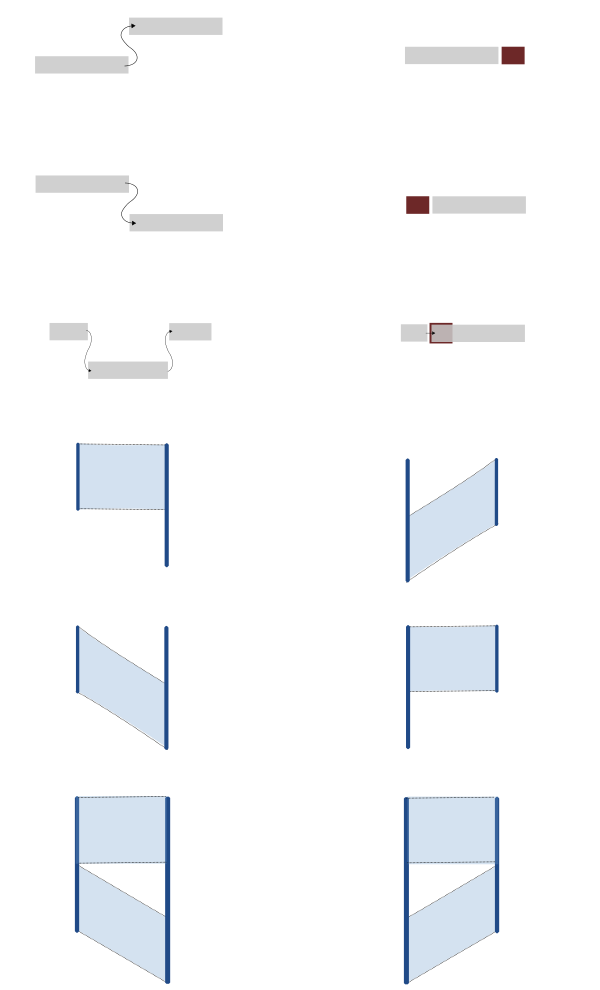}} &
  \subfloat[Delete a part at the beginning of the contents]{\includegraphics[scale=0.60,trim=4.7in 10.10in 0in 1.60in,clip=true]{figure0002.png}}&
  \subfloat[Add at the beginning of the contents]{\includegraphics[scale=0.40,trim=0.50in 3.15in 5.35in 8.65in,clip=true]{figure0002.png}} &
  \subfloat[Delete a part at the beginning of the contents]{\includegraphics[scale=0.40,trim=4.50in 5.50in 1.00in 5.05in,clip=true]{figure0002.png}} \\
  \subfloat[Add at the end of the contents]{\includegraphics[scale=0.60,trim=1in 10.25in 5in 2.4in,clip=true]{figure0002.png}} &
  \subfloat[Delete a part at the ending of the contents]{\includegraphics[scale=0.60,trim=4.7in 12.25in 0in 0.5in,clip=true]{figure0002.png}} &
  \subfloat[Add at the end of the contents]{\includegraphics[scale=0.40,trim=0.50in 5.80in 5.35in 6.15in,clip=true]{figure0002.png}} &
  \subfloat[Delete a part at the end of the contents]{\includegraphics[scale=0.40,trim=4.50in 3.30in 1.00in 8.45in,clip=true]{figure0002.png}} \\
  \subfloat[Add at middle of the contents]{\includegraphics[scale=0.60,trim=1in 8.35in 5in 4.5in,clip=true]{figure0002.png}} &
  \subfloat[Delete a part at middle of the contents]{\includegraphics[scale=0.60,trim=4.7in 8.35in 0in 4.5in,clip=true]{figure0002.png}} &
  \subfloat[Add at middle of the contents]{\includegraphics[scale=0.40,trim=0.50in 0.00in 5.35in 11.05in,clip=true]{figure0002.png}} &
  \subfloat[Delete a part at middle of the contents]{\includegraphics[scale=0.40,trim=4.50in 0.00in 1.00in 11.05in,clip=true]{figure0002.png}} \\
  \end{tabular}
  \caption{Atomic edit operations on CRM (left two column) and on History Flow (right two column). Every atomic operations on CRM is unique while History Flow uses horizontal reflections.} \label{fig:atom}
\end{figure*}

From a purely visual perspective, the fundamental differences between the two approaches are depicted in Figure~\ref{fig:atom}.  Here we represent the two atomic edit operations: add and delete. Note, the CRM scheme represents add persistence through a graph-like structure while HF visualizes the document column-wise.  Perhaps more noteworthy, is the distinction between operations.  CRM atoms have a distinct structure while HF atoms are essentially reflections and/or rotations.  The obvious structural difference in CRM atoms make for quicker more intuitive interpretation.

\section{Evaluation}
We demonstrate our visualization technique using several case studies. These studies include a small size synthetic document, scientific paper written in LaTeX, computer code, and Wikipedia articles. In each of these cases, the revision history was obtained from collaborative tools like Subversion and Wikipedia We focus on the following performance criteria: (a) how easy it is to determine what is the revision in which some change occurs, (b) how easy is it to track what part of the document is changed, (c) which part of the document is frequently edited (d) what was the content that was changed, (e) what was is the style or pattern of the authoring process.

\subsection{Comparison between CRM and History Flow}

In this section, we outline some key differences between CRM and History Flow by visualizing a synthetic document and a Wikipedia article.

\subsubsection{Synthetic Document}

\begin{figure*}
  \centering
  \includegraphics[width=.49\linewidth]{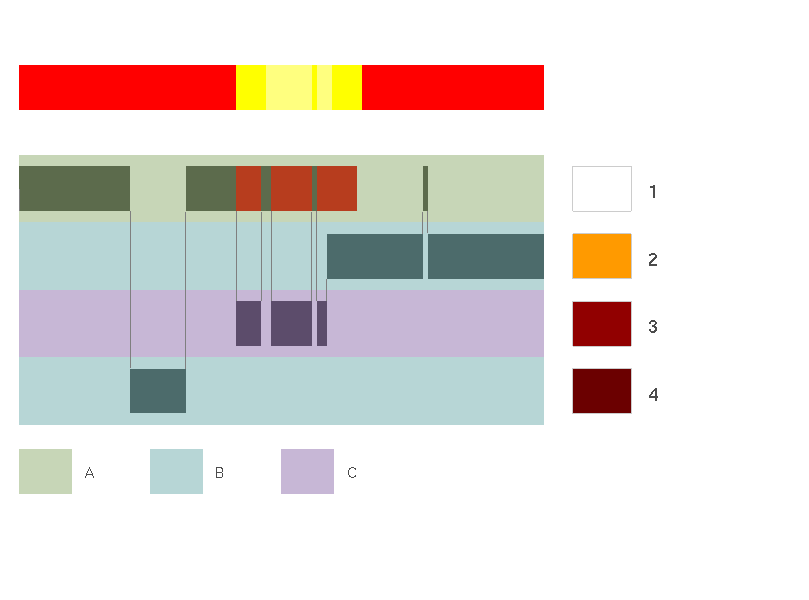}
  \includegraphics[width=.49\linewidth]{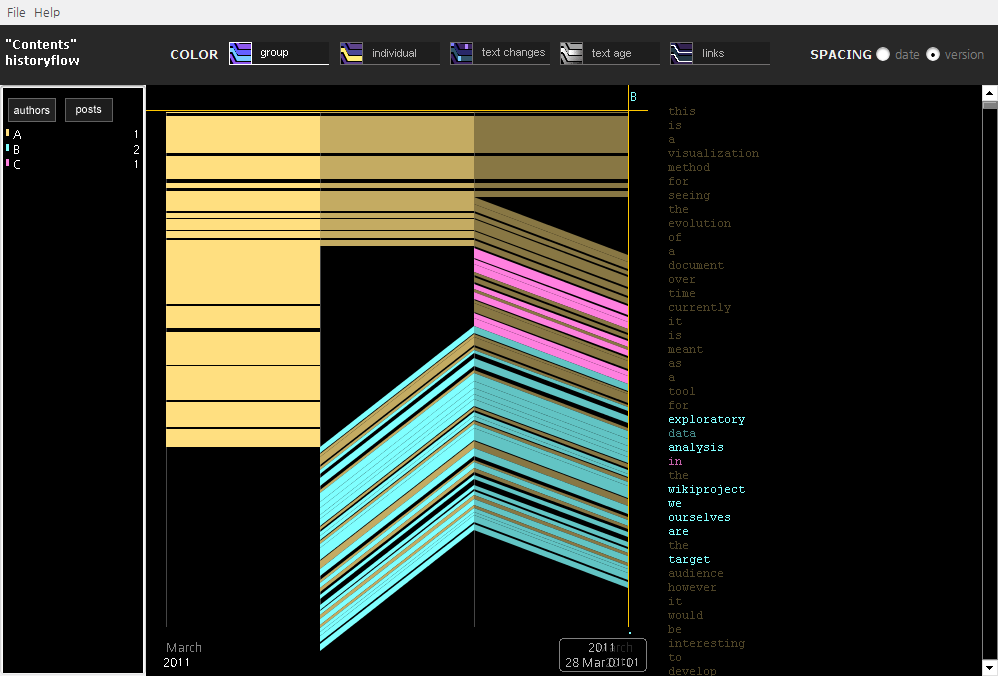}
  \caption{Visualizations of a synthetic document of History Flow website, using CRM(left) and History Flow (right). The document has three authors and four revisions. Gray boxes of CRM(left), linked by vertical lines, indicates persistence contents and red for non-persistent contents. Horizontal band of background color indicates author. Top horizontal bar with spectrum shows degree of cumulative changes in document location: brighter color with higher change. Vertical segmented bar on the right side of CRM means cumulative change in a revision: brighter colors with higher changes. Color bands of History Flow(right) shows the author of relevant content. Horizontal position shows the revision and vertical lengths shows the length of the document at the revision.}  \label{fig:synthetic}
\end{figure*}

Figure~\ref{fig:synthetic} shows a visualization of a short synthetic document using both CRM and the most closely related previous work (History Flow). The synthetic document is the same one that appears in the History Flow website\footnote{\url{http://www.research.ibm.com/visual/projects/history_flow/explanation.htm}}.

The synthetic document has four revision and three authors, and the edit history is the following. User B added content towards the end of the document in the second revision. User C deletes some of the content from the second revision and adds a shorter content instead.  Finally, user B adds some content at around the 25\% percent document position.

CRM (figure~\ref{fig:synthetic}, left) nicely describes the editing patterns. The background colors help to understand which author is active in which revision. The gray boxes indicate persistent changes and red content indicate non-persistent content (removed afterwards in a future revision). Following the gray boxes left to right according to the edges in the figure, we can traverse the final revision of the document and understand which part of that final document was authored by which author and at what time. Specifically, the bottom-left box indicates that there was an addition at revision 4 (bottom row) by user B around the 25\% document position. The boxes at revision 2 (second row, right side) show user B added some contents in the end. Seeing which content persists and which not (red boxes in first row) is very easy. Finding the document positions containing the heaviest editing across revision is easy since the horizontal coordinate indicates the document position and is directly comparable across multiple rows. Moreover, the row on top of the CRM indicates the document positions most heavily edited. 

In contrast, the History Flow visualization (figure~\ref{fig:synthetic}, right)  shows more accurately the relative change of document portions but makes it hard to keep track of document positions across multiple revisions.  The horizontal dimension corresponds to revisions and the vertical dimension corresponds to document position (color shows author).  The expansion and  shrinking of the vertical axis shows the changing length of the document. This may cause confusion as it is hard to compare authoring patterns at different document positions (the horizontal position along the different rows are not comparable). This difficulty of comparing revisions at specific document positions would substantially increase as we have more revisions and more authors. In particular the History Flow visualization of the complete revision history for long documents would exhibit drastic stretching and shrinking which makes it easy to analyze the shift between one revision and the next but makes it hard to detect more global patterns. 

\subsubsection{Wikipedia Article}\label{sec:eval-real}
\begin{figure*}
  \centering
  \includegraphics[width=.85\linewidth]{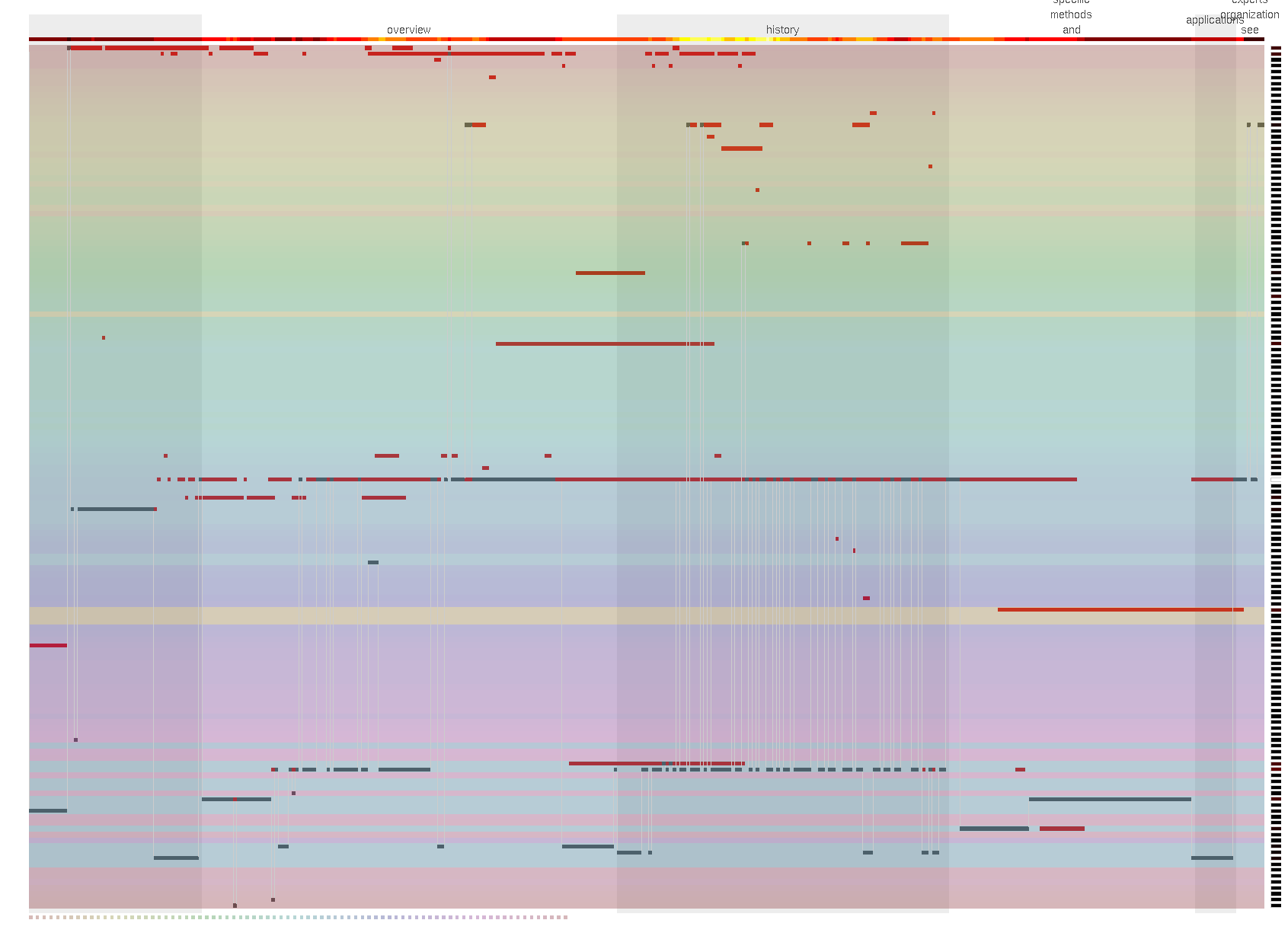}\\
  \includegraphics[width=.85\linewidth]{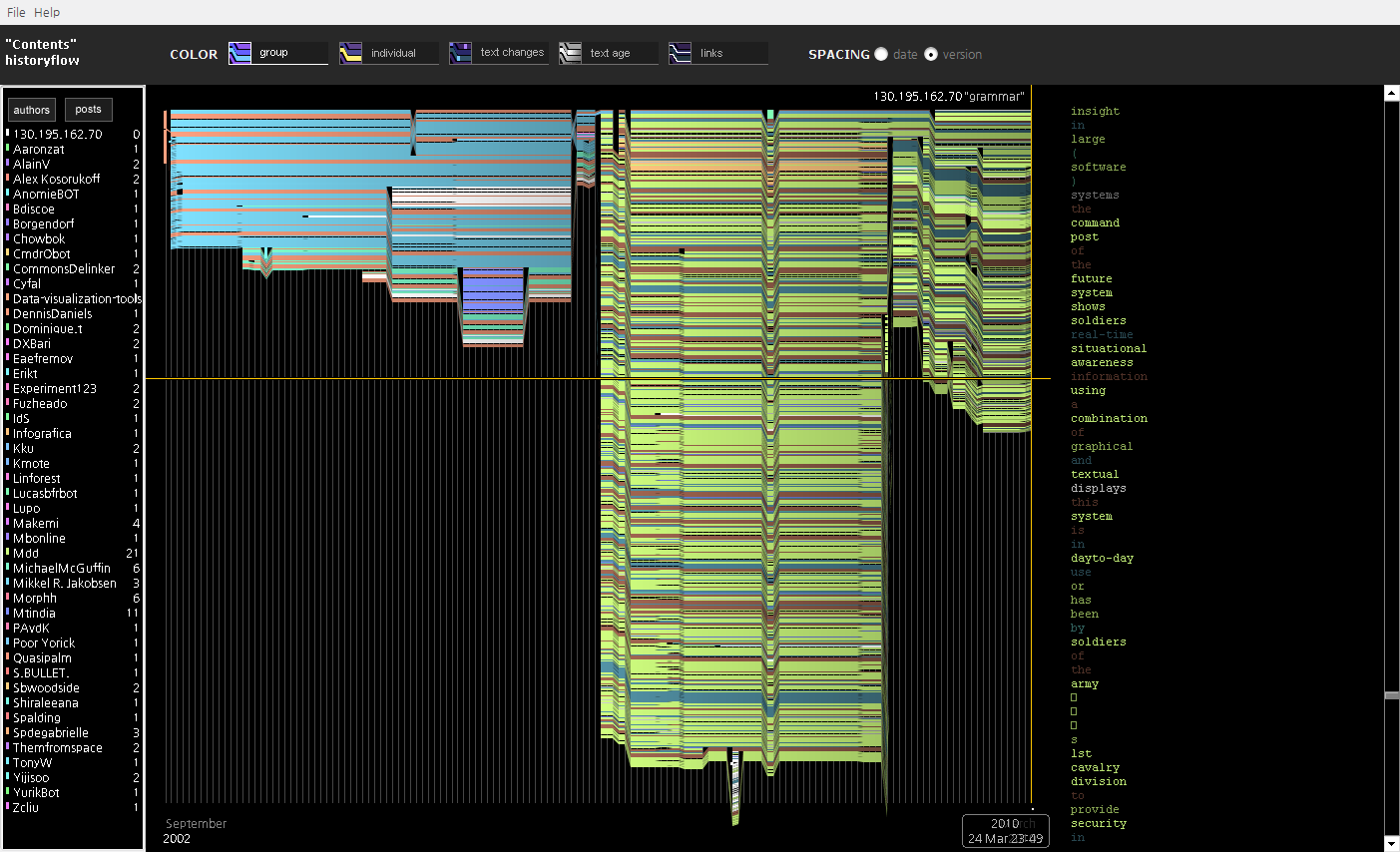}
  \caption{Visualizations of Wikipedia article Information\_Visualization, using CRM(top) and History Flow(bottom). The document has 146 revisions with substantial activity. Some authors wrote content at the beginning that was deleted about halfway through the revision history. Then one user wrote a long version of the article which was substantially trimmed around 80\% of the revision history, after which more content was added to create the present document. Details at Section~\ref{sec:eval-real}. }
  \label{fig:infovis}
\end{figure*}

We turn now to visualizing the authoring process of the Wikipedia article \emph{Information\_Visualization} (146 revisions). This document shows substantial revision activity. For example, some authors wrote content at the beginning that was deleted about halfway through the revision history. Then one user wrote a long version of the article which was substantially trimmed around 80\% of the revision history, after which more content was added to create the present document (March 2011). 

Figure~\ref{fig:infovis} shows the CRM (top) and History Flow visualization of this article. In the CRM case, the upper half of the CRM is almost all in red indicating all content added in the early revisions was later removed. Following left to right along the white edges connecting the gray boxes we can easily track of the revision in which different parts of the final version was authored at. For example, the very first part of the present document comes from about 15 revision before. Almost all gray boxes (persistent changes) are located at the later half of the revision history. The backgrounds at the later half are colored with pink and light blue implying that the present document was authored mostly by two authors. The bar on top of the CRM shows high editing activity (yellow color) on the  \emph{history} section and the middle part of the \emph{overview} section.

The History Flow visualization (bottom figure) shows also that there was a major change in the middle of the revision history and the transition from one group of authors in early revisions to another group in later revisions (indicated by colors changing as we traverse the figure left to right). However, due to extreme shrinking and expanding it is virtually impossible to determine which part was edited in different revisions and how the vertical dimensions relate to each other. As can be seen this is especially difficult for long documents with many active revisions. As a result, it is impossible to determine which parts of the document were most heavily edited throughout the revision history.

\subsection{CRM Case Studies}\label{sec:crm-case}
We demonstrate the CRM visualization on three real world case studies. The case studies show how the CRM may be used to reveal low level and high level collaborative authoring patterns. 

\subsubsection*{LaTeX Conference Paper}
Figure~\ref{fig:crm1} (bottom) shows the CRM of a conference paper written in LaTeX by two of the authors of this paper. The revision history was obtained from a subversion repository. We make the following observations.
\begin{enumerate}
\item There is a striking diagonal editing pattern. This indicates a sequential editing style.  In other words the authors worked their way from the beginning at early revisions towards the middle of the paper in middle revisions to the end of the paper in the final revisions.

\item Author 1 (green horizontal bands) authored a relatively little part of the document (around the middle) and is often being overruled by author 2 (purple horizontal bands). This is indicated by red color which corresponds to edits that are later removed or replaced with other content. This authoring pattern is in agreement with the fact that author 2 is the advisor of author 1 and exhibited a hands on authoring of the paper.  
\end{enumerate}

\subsubsection*{Journal Conference Paper}
\begin{figure*} 
  \centering
  \includegraphics[height=.42\textheight]{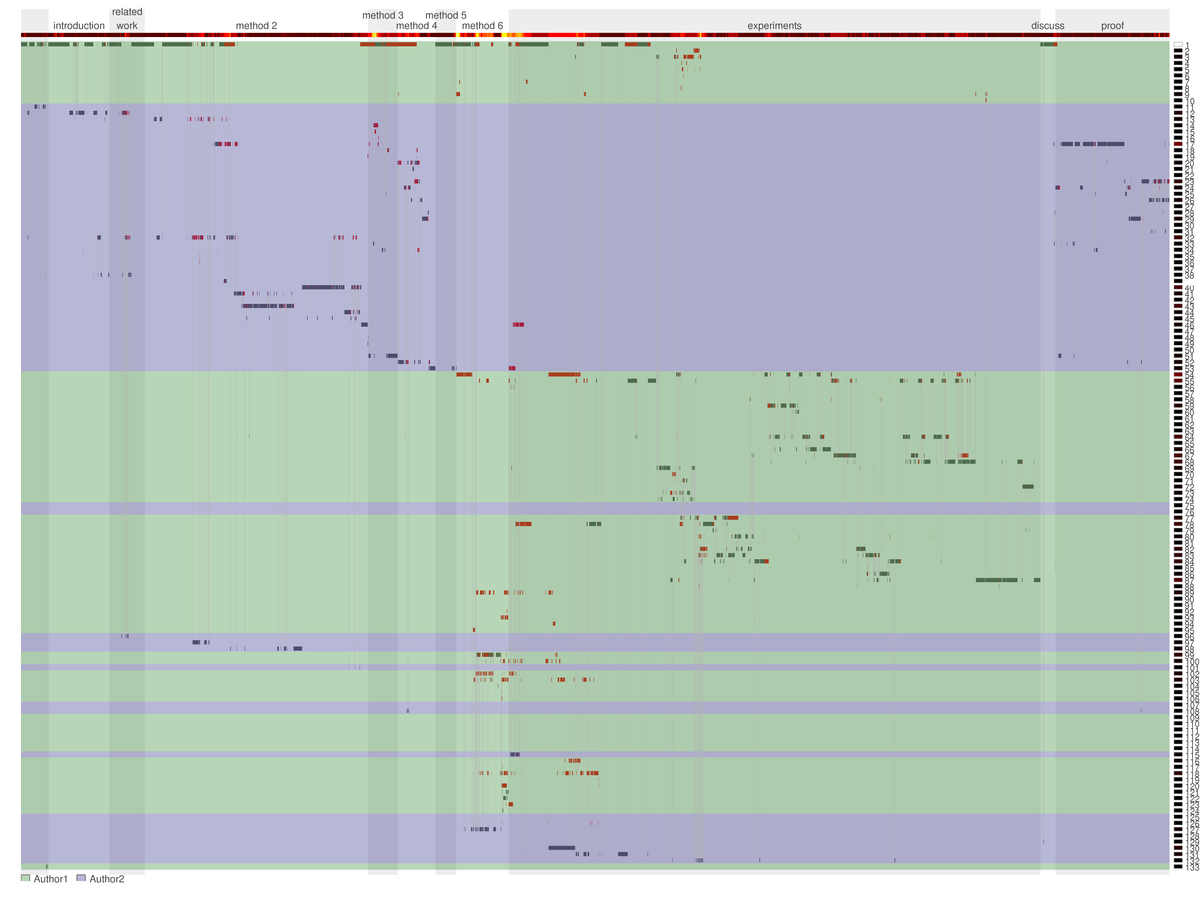}
  \includegraphics[height=.42\textheight]{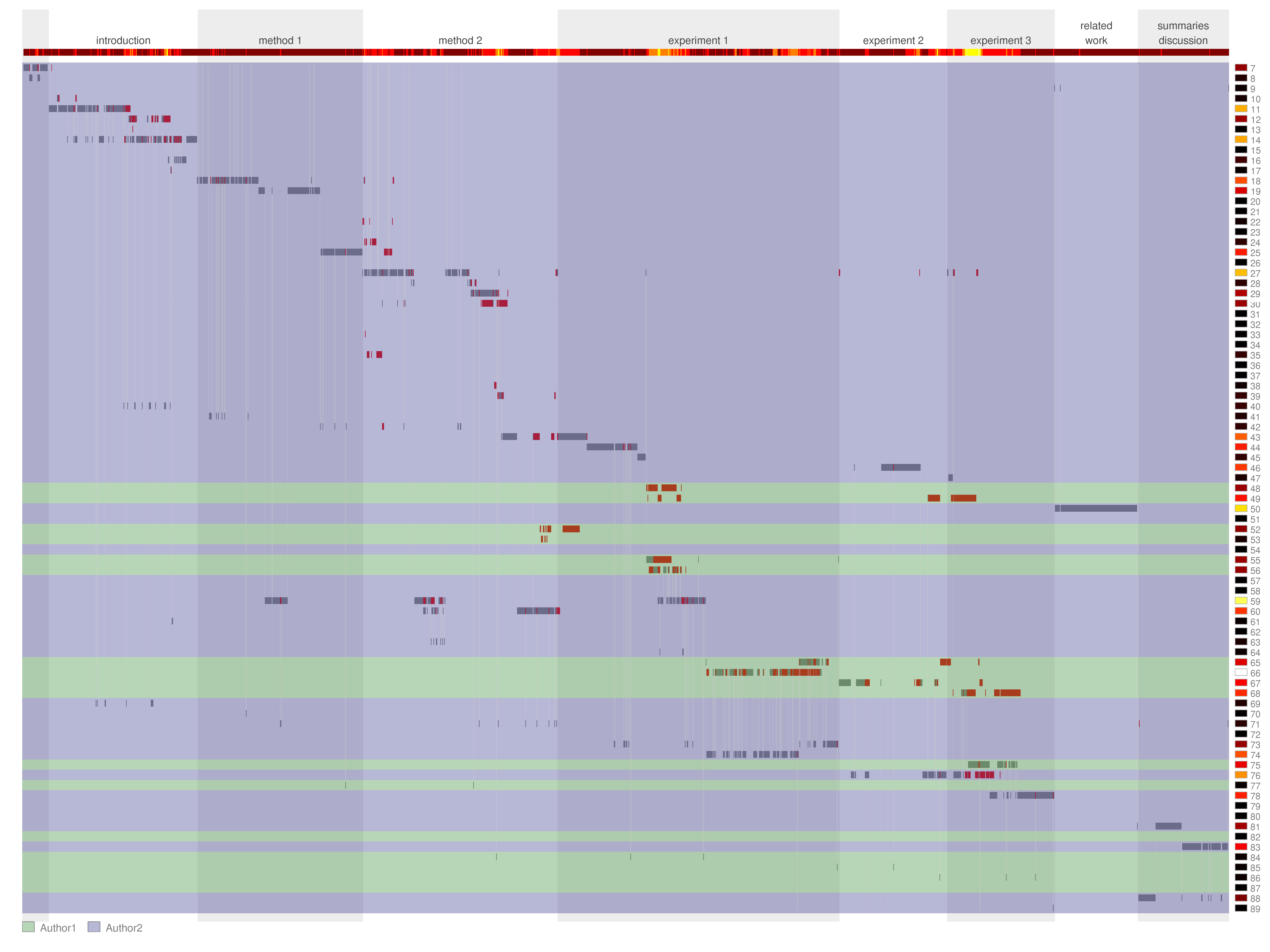}
  \caption{A journal paper(top) and a conference papers(bottom) maintained in SVN. In each CRM, document location flows from left to right, and revisions does top to bottom. Gray boxes represents contents in use in latest revision, and red means deleted contents. Lines are connecting gray boxes along with the content of latest revision. Vertical background bar represents section and horizontal backgrounds shows authors with unique color. Top horizontal bar with spectrum shows degree of cumulative changes in document location: brighter color with higher change. Vertical segmented bar on the right side of each figure means cumulative change in a revision: brighter colors with higher changes. Details at Section~\ref{sec:crm-case}} \label{fig:crm1}
\end{figure*}

Figure~\ref{fig:crm1} (top) shows the CRM of a journal paper written in LaTeX by two of the authors of this paper. The revision history was obtained from a subversion repository. We make the following observations.
\begin{enumerate}
\item  The middle part (method 6 and experiment sections) had the most editing and re-editing by far. On the other hand, the introduction, method, related work, and discussion were not revised much after their initial authoring. 
\item Both authors contribute significantly to the authoring process. Interestingly, the authors do not work much in parallel. Author 1 (green) starts for 10 revisions, author 2 continues for 40 or so revisions, and author 1 resumes the authoring process (with a few exceptions) until the end when both authors make a final pass. The authoring process is very different from the striking diagonal pattern of the conference paper (Figure~\ref{fig:crm1},bottom)
\end{enumerate}

\subsubsection*{Computer Code}
Figure~\ref{fig:crm2} (bottom) shows the CRM of Java computer code from a Google open source project (GWT). We make the following observations.
\begin{enumerate}
\item The code is being repeatedly  overhauled in a significant way. The many red rectangles represent non-persistent changes (edits that do not remain all the way to the final version). Indeed, it seems that the entire first 28 revisions were completely rewritten in the next 20 revisions. 
\item The lack of activity in the beginning of the document (left part does not contain gray or red rectangles) correspond to documentation that is left unchanged. (It is actually a copyright description.)
\item The computer code shows a large deviation from the authoring patterns described in the previous two cases. The computer code was edited in parallel by a large of authors each working on a separate part of the code.
\end{enumerate}

\subsubsection*{Presentation Slides}
\begin{figure*}
  \centering
  \includegraphics[scale=.8, trim=0 0.85cm 0 0, clip=true]{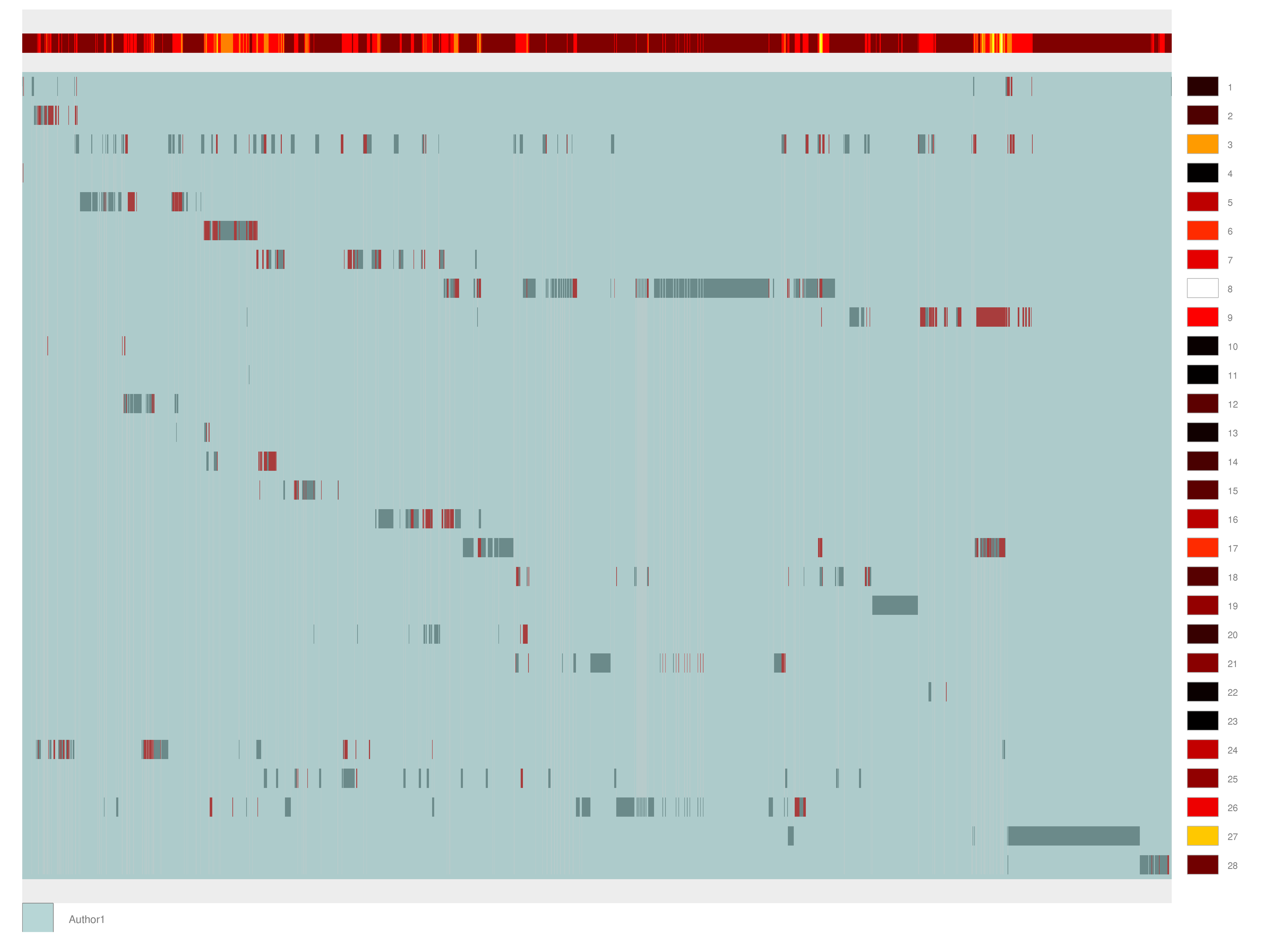}
  \includegraphics[scale=.8, trim=0 0.85cm 0 0, clip=true]{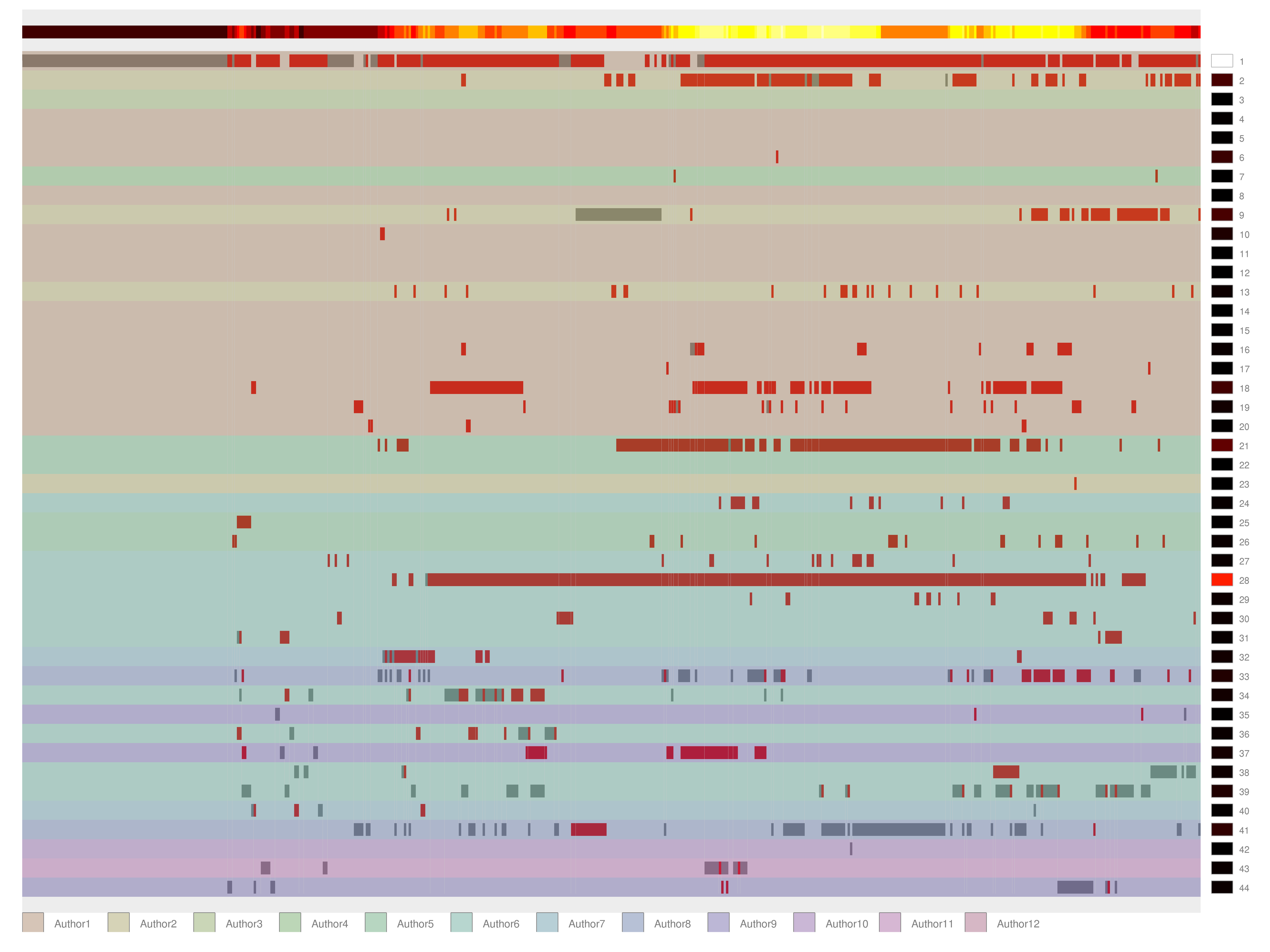}
  \caption{A proposal slide (top), and a Google Tool Kit Compiler source file (bottom). In each CRM, document location flows from left to right, and revisions does top to bottom. Gray boxes represents contents in use in latest revision, and red means deleted contents. Lines are connecting gray boxes along with the content of latest revision. Vertical background bar represents section and horizontal backgrounds shows authors with unique color. Top horizontal bar with spectrum shows degree of cumulative changes in document location: brighter color with higher change. Vertical segmented bar on the right side of each figure means cumulative change in a revision: brighter colors with higher changes. Details at Section~\ref{sec:crm-case}}
  \label{fig:crm2}
\end{figure*}

Figure~\ref{fig:crm2} (top) shows the CRM of LaTeX code corresponding to presentation slides. We make the following observations.
\begin{enumerate}
\item The presentation was authored by a single author who makes four general passes over the slides. Each pass corresponds to a horizontal or nearly horizontal sequence of edits. 
\item Each of the passes show a sequential editing pattern (from start to end) indicated by downward diagonal editing patterns. 
\item The first pass contained relatively light editing (very rough draft) while the other three passes contained more edits. The last pass added a substantial content to the end of the presentation. 
\end{enumerate}

\section{Summary and Discussion}
The past 10 years have seen a substantial increase in the availability and popularity of collaborative authoring tools such as subversion/GIT, MS Office 2010, Google Docs, and Internet wikis and discussion boards. As a result collaborative authoring of documents is becoming more popular and is expected to become even more so in the near future. As the numbers of authors and revisions increase so does the difficulty of understanding the authoring process, both during the authoring stages and in retrospect. Questions such Which author wrote which part? Who removed and edited my contribution? etc., are becoming increasingly hard. 

Thus far most visualization techniques have focused either on visualizing a large corpus of documents or the sequential trends within a single document. An important exception is the History Flow project which is related to our work but focuses on visualizing relative movements of code chunks. Our CRM framework differs by allowing effective visualization of authoring patterns with an emphasis on maintaining a visual relationship between editing patterns and absolute document position. 

The CRM framework can be used to discover low level and high level authoring patterns. Examples of such low level authoring patterns are which sections are edited heavily in which revision, which authors are more active than others, and what parts of the document are re-edited multiple times. Examples of high level authoring patterns are sequential vs parallel authoring, which authors re-edit the text of other authors, and identifying large structural changes such as section rearrangements. 

\clearpage
\bibliographystyle{abbrv}
\bibliography{../../share/externalPapers,../../share/groupPapers}

\begin{thebibliography}{10}

\bibitem{Yates1999}
R.~Baeza-Yates and B.~Ribeiro-Neto.
\newblock {\em Modern Information Retrieval}.
\newblock Addison Wesley, 1999.

\bibitem{Blei2006}
D.~Blei and J.~Lafferty.
\newblock Dynamic topic models.
\newblock In {\em Proc. of the International Conference on Machine Learning},
  2006.

\bibitem{Cleveland1993}
W.~S. Cleveland.
\newblock {\em Visualizing Data}.
\newblock Hobart Press, 1993.

\bibitem{Oliveira2003}
M.~F. de~Oliveira and H.~Levkowitz.
\newblock From visual data exploration to visual data mining: A survey.
\newblock {\em IEEE Transactions on Visualization and Computer Graphics}, 9(3),
  2003.

\bibitem{Diehl2007}
S.~Diehl.
\newblock {\em Software Visualization: Visualizing the Structure, Behaviour,
  and Evolution of Software}.
\newblock Springer, 2007.

\bibitem{Eick2002}
S.~G. Eick, T.~L. Graves, A.~F. Karr, A.~Mockus, and P.~Schuster.
\newblock Visualizing software changes.
\newblock {\em Transactions on Software Engineering}, 28(4):396--412, 2002.

\bibitem{Eick1992}
S.~G. Eick, J.~L. Steffen, and J.~E.~E.~Sumner.
\newblock Seesoft--a tool for visualizing line oriented software statistics.
\newblock {\em Transactions on Software Engineering}, 18(11):957--968, 1992.

\bibitem{Chevalier2010}
A.~B. F.~Chevalier, P.~Dragicevic and J.~Fekete.
\newblock Using text animated transitions to support navigation in document
  histories.
\newblock In {\em Proceedings of the 28th international conference on Human
  factors in computing systems}, CHI '10, pages 683--692. ACM, 2010.

\bibitem{Fodor2002}
I.~K. Fodor.
\newblock A survey of dimension reduction techniques.
\newblock Technical Report UCRL-ID-148494, Lawrence Livermore National
  Laboratory, 2002.

\bibitem{Fortuna2005}
B.~Fortuna, M.~Grobelnik, and D.~Mladenic.
\newblock Visualization of text document corpus.
\newblock {\em Informatica}, 29:497--502, 2005.

\bibitem{Froehlich2004}
J.~Froehlich and P.~Dourish.
\newblock Unifying artifacts and activities in a visual tool for distributed
  software development teams.
\newblock {\em International Conference on Software Engineering}, pages
  387--396, 2004.

\bibitem{Fry2000}
B.~J. Fry.
\newblock Organic information design.
\newblock Master's thesis, School of Architecture and Planning, MIT, 2000.

\bibitem{Havre2001}
S.~Havre, E.~Hetzler, K.~Perrine, E.~Jurrus, and N.~Miller.
\newblock Interactive visualization of multiple query results.
\newblock In {\em IEEE Symposium on Information Visualization}, pages 105--112,
  2001.

\bibitem{Havre2002}
S.~Havre, E.~Hetzler, P.~Whitney, and L.~Nowell.
\newblock Themeriver: Visualizing thematic changes in large document
  collections.
\newblock {\em IEEE Transactions on Visualization and Computer Graphics},
  8(1):9--20, 2002.

\bibitem{Hochheiser2004}
H.~Hochheiser and B.~Shneiderman.
\newblock Dynamic query tools for time series data sets, timebox widgets for
  interactive exploration.
\newblock {\em Information Visualization}, 3(1):1--18, 2004.

\bibitem{Jelinek1999}
F.~Jelinek.
\newblock {\em Statistical methods for speech recognition}.
\newblock MIT press, 1999.

\bibitem{Lee2007}
J.~Lee and M.~Verleysen.
\newblock {\em Nonlinear dimensionality reduction.}
\newblock Springer, 2007.

\bibitem{Manning1999}
C.~D. Manning and H.~Schutze.
\newblock {\em Foundations of Statistical Natural Language Processing}.
\newblock MIT Press, 1999.

\bibitem{Ogawa2008}
M.~Ogawa and K.-L. Ma.
\newblock Stargate: A unified, interactive visualization of software projects.
\newblock {\em In Proceedings of the 2008 IEEE Pacific Visualization
  Symposium}, pages 191--198, 2008.

\bibitem{Ogawa2009}
M.~Ogawa and K.-L. Ma.
\newblock code swarm: A design study in organic software visualization.
\newblock {\em IEEE Transactions on Visualization and Computer Graphics},
  15(6):1097--1104, 2009.

\bibitem{Spoerri1993}
A.~Spoerri.
\newblock {InfoCrystal}: A visual tool for information retrieval.
\newblock In {\em Proceedings of the 4th Conference on Visualization}, 1993.

\bibitem{Storey2005}
M.-A.~D. Storey, D.~Cubrani, and D.~M. German.
\newblock On the use of visualization to support awareness of human activities
  in software development: a survey and a framework.
\newblock {\em SOFTVIS}, pages 192--201, 2005.

\bibitem{Thomas2005}
J.~J. Thomas and K.~A. Cook, editors.
\newblock {\em Illuminating the Path}.
\newblock IEEE Computer Society, 2005.

\bibitem{nvacbook}
J.~J. Thomas and K.~A. Cook, editors.
\newblock {\em Illuminating the Path: The Research and Development Agenda for
  Visual Analytics}.
\newblock IEEE Computer Society, 2005.

\bibitem{Tufte2001}
E.~R. Tufte.
\newblock {\em The Visual Display of Quantitative Information}.
\newblock Graphic Press, 2 edition, 2001.

\bibitem{Viegas2006}
F.~B. Vi\'{e}gas, S.~Golder, and J.~Donath.
\newblock Visualizing email content: portraying relationships from
  conversational histories.
\newblock In {\em Proc. of the conference on Human Factors in computing
  systems, {CHI}'06}, pages 979--988, 2006.

\bibitem{Viegas2004}
F.~B. Vi\'{e}gas., M.~Wattenberg, and K.~Dave.
\newblock Studying cooperation and conflict between authors with history flow
  visualizations.
\newblock In {\em Proceedings of the SIGCHI conference on Human factors in
  computing systems}, CHI '04, pages 575--582. ACM, 2004.

\bibitem{Weber2001}
M.~Weber, M.~Alexa, and W.~Muller.
\newblock Visualizing time-series on spirals.
\newblock In {\em Proc. of IEEE Symposium on Information Visualization}, pages
  7--14, 2001.

\end{thebibliography}

\end{document}